\documentstyle[12pt]{article}

\begin{document}

\begin{titlepage}
\begin{flushright}
\hfil                 UCSD/PTH 92-02\\
\hfil                 BNL-45082\\
\hfil Jan. 1992 \\
\end{flushright}
\begin{center}

{\LARGE\bf Tunneling and Energy Splitting in Ising Models}

\vspace{1.5cm}
\large{
Karl Jansen\\
Department of Physics, University of California at San Diego,\\
La Jolla, CA~92093-0319, USA  \\
\vspace{0.5cm}
and\\
\vspace{0.3cm}
Yue Shen$^{1}$\\
Physics Department, Brookhaven National Laboratory\\
Upton, NY 11973, USA \\

}
\vspace{1.9cm}
\end{center}

\abstract{
The energy splitting $E_{0a}$ in two and four dimensional Ising models is
measured in a cylindrical geometry on finite lattices.
By comparing to exact results in the two dimensional
Ising model we demonstrate that $E_{0a}$ can be extracted very
reliably from Monte Carlo calculations in practice.
In four dimensions we compare the measured $E_{0a}$ with
two different theoretical predictions on the finite size behavior of the
energy splitting.  We find that our numerical data are in favor of the
predictions based on the semiclassical dilute instanton gas approximation.
}
\vfill
\footnotetext[1]{{\em  Research supported under contract number
DE-AC02-76CH00016 with the U.S. Department of Eneregy. Accordingly,
the U.S. Government retains a non-exclusive, royalty-free license
to publish or reproduce the published form of this contribution, or
allow others to do so, for U.S. Government purposes.
}}
\end{titlepage}

\section{Introduction}

In studies of first order phase transitions
on finite lattices one faces the phenomenon of
tunneling between different states.
The tunneling rate is related to the surface tension, which is the free
energy per unit area associated with the interface
separating these different states. As the surface tension is the
fundamental parameter for the nucleation rate in first order phase
transitions the determination of it has attracted a lot of attention.
Examples are widespread. In d=2, models like the $q$-state potts model
are of interest for statistical physics. In d=3 binary
liquids give direct experimental information about the surface tension
that can be compared to theoretical calculations
using the Ising model as an effective description. In d=4 it is the
first order QCD
finite temperature phase transition where attempts have been made to extract
the surface tension.
This list of examples is of course by far not exhausting and many more
examples could be found.

Here we study the two and four dimensional Ising models below the critical
temperature as examples of a typical system undergoing a first order phase
transition between two ordered states. In a cylindrical geometry where
the time dimension is much longer than the linear spatial dimension, the
Ising system will break into domains along the time direction through
tunneling. The inverse of the correlation length along the time direction
defines a energy splitting between the ground state and the first excited
state.
This energy splitting is related to the surface tension \cite {fisher} which,
in turn, is associated with the domain wall energy. The finite volume
dependence of the energy splitting has been a long standing problem. A few
years
ago, a definite finite size behavior for the energy splitting was derived
based on a semiclassical dilute instanton gas approximation \cite {munster}.
The result agreed with an earlier intuitive conjecture by Br{\' e}zin
and Zinn-Justin \cite {brezin}.
Furthermore, a relation between the surface tension
and the renormalized parameters was derived in this approximation.
This relation was checked in a numerical simulation and good
agreement had been found both in four \cite {jansen} and three
dimensions \cite{mot}.

Recently, Borgs and Imbrie \cite {borgs} gave a new derivation for the
finite size behavior of the energy splitting for a general Ising like system.
Their result is claimed to be {\it exact}. However,
it is in disagreement with the instanton calculation.

In this paper we confront both theoretical predictions with a Monte
Carlo simulation. We first calculate the energy splitting in the two
dimensional
Ising model. By comparing to the known exact formula, we demonstrate that
the method developed in ref. \cite {jansen} is very efficient and reliable
for calculating the energy splitting. Then we turn our attentions to the four
dimensional Ising model.

\pagebreak
\section{The Ising model}

The Ising model in d-dimensions is defined by

\begin{equation}
S = -\kappa\sum_{x\in\Lambda}\sum_{\mu=1}^d \Phi_x\left[\Phi_{x+\mu}
+ \Phi_{x-\mu} \right]~,
\label{eq:model}
\end{equation}
\noindent where the fields $\Phi$ can only assume values $\pm 1$. Here
$\Lambda$ is the volume of the lattice, i.e. $TL$ in d=2 and $TL^3$ in
d=4 with $T$ the time and $L$ the spatial extent of the lattice.
Periodic boundary conditions are taken in all directions.
The hopping parameter $\kappa$ is related to
the inverse temperature. As is well known, the model
eq. (\ref{eq:model}) can be solved exactly in d=2 \cite{onsager} and
has therefore played the role as a very often used toy model to test ideas and
methods. In d=4 the Ising model is in the same universality class as the
$\Phi^4$ theory. It exhibits therefore properties like triviality
and was used to study at least qualitatively important aspects of
the Standard Model physics \cite{kushe,luwe,jansen}.

The Ising model eq. (\ref{eq:model}) is symmetric under parity
transformation $\Phi \to - \Phi$.
For $\kappa > \kappa_c$ with $\kappa_c=\frac{1}{4}\log{(1+\sqrt{2})}
\approx 0.22036$ in d=2 and $\kappa_c\approx 0.0748$ in d=4, this
global Z(2) symmetry is spontaneously broken in infinite volume.
We have a double well potential with minima at
$+M$ and $-M$. There are two degenerate ground states $|0_+>$ and $|0_->$
which can be intuitively thought of as wavepackets
centered around $+M$ and $-M$,
respectively. $|0_+>$ and $|0_->$ have no overlap and a complete Hilbert
space for the system can be built on either one of these ground states.
Assuming we choose $|0_+>$ as the ground state, the system will have a
nonzero magnetization $<0_+|\Phi|0_+> = M$.
If one includes an external source $j$ the magnetization will align
with $j$. With varying external source
the system will pass a first oder phase transition at
$j=0$ and the magnetization $M$ jumps from $+M$ to $-M$.

In a finite system the parity symmetry is never spontaneously broken. For
$\kappa > \kappa_c$ the $|0_+>$ and $|0_->$ states are no longer the ground
states and have finite overlap.
A wavepacket centered around one minimum of the effective potential
can always tunnel and spread over to the other minimum. The ground state
$|0_s>$ is symmetric under parity transformation and may be expressed
as
\begin{equation}
|0_s>  =  \frac{1}{\sqrt{2}}(|0_+> + |0_->) .
\label{eq:even}
\end{equation}
The antisymmetric linear combination of $|0_+>$ and $|0_->$ forms
the parity odd first excited state $|0_a>$
\begin{equation}
|0_a>  =  \frac{1}{\sqrt{2}}(|0_+> - |0_->).
\label{eq:odd}
\end{equation}
If we set the ground state energy to be zero, there is a small energy splitting
$E_{0a}$ between the first excited state $|0_a>$ and the ground state
$|0_s>$. This energy splitting is due to the
tunneling between $|0_+>$ and $|0_->$ states.

As is well known \cite{fisher,privman,coleman} the energy splitting $E_{0a}$
vanishes exponentially fast with the spatial extent of the lattice and
is related to the surface tension $\sigma$,
$E_{0a} \sim exp(-\sigma L^{d-1})$.
The finite size behavior of $E_{0a}$ can be analyzed by choosing a
lattice geometry where the time extent $T$ is much larger than the
spatial extent $L$ of the lattice. In this so called cylindrical geometry
the system breaks up into domains of different magnetization, separated
by interfaces in the spatial direction.

Defining the timeslice operator $S_t$
\begin{equation}
S_t = \sum_{\tilde{x}\in L^{d-1}} \Phi_{\tilde{x},t}~,
\label{operator}
\end{equation}
for large enough $t$
the tunneling energy can be obtained from the correlation function
\begin{equation}
<S_0S_t> = c_1(e^{-E_{0a}t} + e^{-E_{0a}(T-t)})\; .
\label{eq:correlat}
\end{equation}

We calculated the correlation function $<S_0S_t>$ numerically in the d=2
and d=4 Ising models in the broken phase. We used the cluster algorithm
\cite{swendsen,wolff}
in its 1-cluster version where we performed 10 updates in between
measurements. We used both, the correlation function $<S_0S_t>$ and its
improved estimator version \cite {sweeny}
\begin{equation}
<S_0S_t> = <\frac{\Lambda}{|c|}\sum_{{\tilde x},{\tilde y}}\Theta({\tilde x},t;
{\tilde y},0) >\; .
\label{eq:improved}
\end{equation}
\noindent Here $|c|$ is the size of the cluster $C$ and $\Theta$ is the
cluster incidence function
\begin{equation}
\Theta(x;y) = \left\{ \begin{array}{ll}
                      1 & \mbox{if both $x,y \in C$} \\
                      0 & \mbox{else.}
                      \end{array} \right.
\end{equation}

To achieve a realistic error for the energies $E_{0a}$ we split the runs into
several subruns and fitted the data according to eq. (\ref{eq:correlat})
for each subrun. We then determined the values of $E_{0a}$ by the mean
and its error of these individual results.

\section{The two dimensional Ising model}

We have chosen the two dimensional Ising model to test the method of
extracting the surface tension from the finite size behavior of
$E_{0a}$. In two dimensions the $L$-dependence of $E_{0a}$ in the
cylindrical geometry is known exactly \cite{privman,onsager}
\begin{equation}
E_{0a}=\frac{1}{2}\sum_{k=0}^{L-1}
\left[\gamma\left(\frac{2k+1}{L}\right) -\gamma\left(\frac{2k}{L}\right)\right]
\label{eq:exact1}
\end{equation}
where
\begin{equation}
\gamma(x)=\cosh^{-1}\left[\cosh(2\kappa^\ast)\cosh(4\kappa)
         -                \cos(\pi x)
                    \right]
\label{eq:gamma}
\end{equation}
\noindent and $\kappa^\ast$ is the dual coupling
\begin{equation}
\tanh(\kappa^\ast) = e^{-4\kappa} \; .
\label{eq:dual}
\end{equation}

This exact expression has an asymptotic expansion for $L \rightarrow
\infty$
\begin{equation}
E_{0a}=
       \sqrt{\frac{2\sinh(\sigma)}{\pi L}}e^{-L\sigma}(1+O(\frac{1}{L})) ~,
\label{eq:asymptotic}
\end{equation}
\noindent where the surface tension $\sigma$ is given by
\begin{equation}
\sigma = 2(2\kappa -\kappa^\ast) = 4\kappa - \ln \left[{1 + e^{-4\kappa} \over
1 - e^{-4\kappa} } \right]\; .
\label{exactsurface}
\end{equation}

Our lattices had a geometry $LT$ where $T=1000$ and $L$ varied
from $5-16$. We have chosen two $\kappa$ values
in the broken phase of the model, $\kappa =0.2326$ and
$\kappa=0.2252$. We performed
50,000 measurements with 10 cluster updates in between each
measurement. We fitted the spin-spin correlation function
eq. (\ref{eq:correlat}) and the improved estimator eq. (\ref{eq:improved})
according to
eq. (\ref{eq:correlat}) and performed a block error analysis as described
above. Both operators gave consistent results. We checked that our fit
results are stable against varying the start and end point of the fits.

The resulting values of $E_{0a}$ together with the exact
formula eq. (\ref{eq:exact1}) is displayed in Fig. 1. We find a perfect
agreement between the numerical data and the exact formula. This shows
that the energy splitting $E_{0a}$ can be extracted from the correlation
function very precisely and reliably.

We remark here that although $E_{0a}$ can be measured accurately,
extracting the surface tension from $E_{0a}$ is not easy in this case.
The surface
tension is only clearly defined when $L$ is large enough and $E_{0a}$
can be well approximated by its asymptotic form eq. (\ref{eq:asymptotic}).
In $d=2$ eq. (\ref{eq:asymptotic}) sets in very slowly. For
example, at $\kappa=0.2326$, $E_{0a}$ calculated from eq. (\ref{eq:asymptotic})
is $20\%$ smaller than the exact value even when $L=16$. In higher dimensions
we expect the situation to improve since the asymptotic behavior of
$E_{0a}$ should set in much earlier with an error of $O(1/L^{d-1})$.

\section{The four dimensional Ising model}

It is well known that the Ising model is in the same universality class
as the $\lambda\Phi^4$ theory. Using the language of the continuous
$\lambda\Phi^4$ field the energy splitting
$E_{0a}$ can be calculated in a dilute
instanton gas approximation \cite{coleman}. If we write the bare potential
as
\begin{equation}
V(\Phi) = - {m_0^2 \over 4} \Phi^2 + {g_0 \over 24} \Phi^4 ~,
\end{equation}
the energy splitting can be written as
\begin{equation}
E_{0a} = 2 e^{-S_c} \left({S_c \over 2\pi}\right)^{1/2} \left | {det^\prime M
\over det M_0} \right |^{-1/2} ~,
\label{eq:E0abare}
\end{equation}
where $det^\prime$ is the determinant without the zero mode and
the classical action $S_c$ comes from a single instanton configuration
\begin{equation}
S_c = 2\frac{m_0^3}{g_0}L^{d-1} \; ,
\label{eq:classaction}
\end{equation}
and the determinants $M$ and $M_0$ are given by
\begin{equation}
M = -\partial^2 + m_0^2 - {3\over 2}m_0^2 cosh^{-2}\left({m_0\over 2}x_4\right)
{}~,
\end{equation}
and
\begin{equation}
M_0 = -\partial^2 + m_0^2~.
\end{equation}
In eq. (\ref {eq:E0abare}) various factors can be understood intuitively.
The factor $\exp (-S_c)$ comes from summing over a
noninteracting multi-instanton configuration, the $\sqrt{S_c}$ factor
comes from an integration over the location of the instantons and the ratio
of determinants $det^\prime M/ det M_0$ comes from integrating over
the Gaussian fluctuations around the classical instanton configuration.
The classical instanton action eq. (\ref{eq:classaction}) scales with the
spatial volume $L^{d-1}$.
The ultraviolet part in the determinant ratio $|det^\prime M/ det
M_0|^{-1/2}$ will renormalize the bare parameters in $S_c$ and the infrared
part
was conjectured to contribute an additional factor $L^{-1}$ \cite {brezin}.
Thus the energy splitting has a general form
\begin{equation}
E_{0a} = C_{inst}L^{\omega}e^{-\sigma L^{d-1}}~,
\label{eq:E0a}
\end{equation}
where $\sigma$ is the surface tension and $\omega$ is conjectured to be
$(d-3)/2$ under the dilute instanton gas approximation.

M{\" u}nster has calculated the determinant ratio $det^\prime M /
det M_0$ in a finite spatial volume \cite {munster}.
The result of his work is in agreement with the conjectured form
eq. (\ref{eq:E0a}) with $\omega= (d-3)/2$. In addition the surface tension and
the constant $C_{inst}$ were calculated
in terms of the renormalized parameters.
For example, in $d=4$, we have
\begin{equation}
\sigma_{inst} = 2\frac{m_r^3}{g_r}
\left[1-\frac{g_r}{16\pi^2}\left(\frac{1}{8}+\frac{\pi}{4\sqrt{3}}\right)
+ O(g_r^2)\right]
\label{eq:sigmainstanton}
\end{equation}
\noindent and
\begin{equation}
C_{inst}=1.65058\sqrt{2\frac{m_r^3}{g_r}}~.
\label{eq:cinstanton}
\end{equation}
\noindent In eqs. (\ref{eq:sigmainstanton}) and (\ref{eq:cinstanton}) $m_r$
and $g_r$ are the renormalized mass and renormalized quartic coupling
respectively. These quantities are input parameters which have to be
determined. A high statistic Monte Carlo simulation was performed
at $\kappa = 0.076$ \cite{jansen}. $E_{0a}$, $m_r$ and $g_r$ were measured
independently and the relation given by eq. (\ref{eq:sigmainstanton}) was
confirmed.

Recently, Borgs and Imbrie claimed
\cite{borgs} to have calculated the {\em exact} finite
size behavior of the energy splitting. In $d=4$ they get
\begin{equation}
E_{0a} = 2e^{-\sigma L^3}~, \ \ \   \kappa > \kappa_c~.
\label{eq:bie0a}
\end{equation}
This formula  differs from
eq. (\ref{eq:E0a}) in two aspects. First, the prefactor in
front of $\exp (-\sigma L^3)$ is independent of the volume.
Second, the numerical value of the pre-exponential factor equals
two and is believed to be independent of $\kappa$.

In order to confront both theoretical predictions with numerical
simulation results we performed Monte Carlo simulations to determine
$E_{0a}$.
In the same manner as in d=2 we studied the correlation function
eq. (\ref{eq:correlat}) and its improved estimator eq. ({\ref{eq:improved})
also in the four dimensional case. Here we have chosen four $\kappa$ values
where the results for $\kappa =0.076$ are taken from reference \cite{jansen}.
The measured $E_{0a}$ values are listed in Table 1.
As mentioned above the values of $g_r$ and $m_r$ are needed as input for
the coefficient $C_{inst}$ and the surface tension $\sigma_{inst}$
in the instanton calculation.
For the $\kappa$ values (beside $\kappa=0.0760$) used here
we measured $m_r$ and $\lambda_r$ on large lattices in a similar way as in
\cite{jansen}. The resulting values of $m_r$ and $g_r$ can be found in
Table 2.

We show the finite size behavior of $E_{0a}$ for all four $\kappa$
values in Fig. 2. The solid lines are the {\em predictions} from the
instanton calculation eqs. (\ref{eq:E0a}),
(\ref{eq:sigmainstanton}) and (\ref{eq:cinstanton}) and the solid circles
are our measured $E_{0a}$. One of the
assumptions in the instanton calculation is that the linear size of the
spatial volume $L$ is large. Thus we expect the agreement with
MC data will only set in when $L > 2 \xi$ with $\xi=1/m_r$ being the
correlation
length in a hypercubic geometry. This seems to be the case as
shown in Fig. 2. Our largest spatial volume is $L=14$ at $\kappa=0.0751$
which is not obviously large enough so that the asymptotic behavior for
$E_{0a}$ has set in. At $\kappa=0.0754$ and $L=12$
we checked that there is $18\%$ finite size effect when $T$ is increased
from $120$ to $160$
and the effect is reduced to $6\%$ at $L=11$.
This means that $T$ should be at least
more than 10 times larger than $L$ to suppress the finite size effects
in time direction. Thus an increase in $L$ would require a large
increase in $T$
which will require a substantial increase in computer time. Given our limited
numerical data, it is not possible to determine the power $\omega$ of $L$
dependence in the pre-exponential factor in eq. (\ref{eq:E0a}).
Instead, at each $\kappa$ value we take 3 data points with the
largest $L$ and fit to the form
\begin{equation}
\ln {E_{0a} \over \sqrt{L}} = - \sigma_{fit} L^3 + \ln C_{fit} ~.
\label{eq:fit1}
\end{equation}
The fitted results for $\sigma_{fit}$ and $C_{fit}$ are listed in Table 3,
along
with $\chi^2$ per degrees of freedom. For comparison we also list in Table 3
the predicted $\sigma_{inst}$ and $C_{inst}$ values from
eqs. (\ref{eq:sigmainstanton}) and (\ref{eq:cinstanton}).
It is clear from Table 3 that the measured $\sigma_{fit}$ and $C_{fit}$ values
are in reasonable agreement with the instanton calculation prediction.
Given the fact that the one-loop correction term in eq.
(\ref{eq:sigmainstanton})
contributes from $9\%$ at $\kappa=0.0751$ to $13\%$ at $\kappa=0.0764$,
it is unlikely that the observed agreement with the fitted
numerical results is accidental.

We also tried a fit to the form
\begin{equation}
E_{0a} = Ce^{-\sigma L^3}
\label{eq:fit2}
\end{equation}
and the results are listed in Table 4. From $\chi^2$ both fits
eq. (\ref{eq:fit1}) and eq. (\ref{eq:fit2}) are of the
same quality with a slightly smaller $\chi^2$ for eq. (\ref{eq:fit2}).
The fitted values for $\sigma$ are not very far from those listed in Table 3.
However, the fitted pre-exponential factor
$C$ values are much smaller than 2 and are
dependent on the value of $\kappa$.
This is a contradiction to eq. (\ref{eq:bie0a}).

A direct comparison to Borgs and Imbrie result
is not possible because in eq. (\ref{eq:bie0a}) the surface tension $\sigma$
is not expressed in terms of the renormalized parameters and should be
treated as a free parameter.
However, since both fits to eq. (\ref{eq:fit1}) and eq. (\ref{eq:fit2})
give approximately the same values for the surface tension as shown
in Tables 3 and 4 and those values are in
reasonable agreement with the predicted value $\sigma_{inst}$ given by
eq. (\ref{eq:sigmainstanton}), we may take $\sigma_{inst}$ value as
an estimate of the true surface tension.
If we use the value of $\sigma_{inst}$ in eq. (\ref{eq:bie0a}) and setting the
coefficient to 2 as predicted by Borgs and Imbrie,
the calculated $E_{0a}$ will have a huge difference from MC data as shown
in Fig. 3 ($\kappa = 0.0754$).

\section{Conclusion}

Choosing a cylindrical geometry where the time extent of the lattice is
much larger than the spatial extent, we studied the finite size
behavior of
the energy splitting $E_{0a}$ on periodic lattices. We demonstrated in the two
dimensional Ising model that the values of $E_{0a}$ can be
extracted from the spin-spin correlation function very precisely.
The finite size dependence of $E_{0a}$ was compared to the known exact results.
We find perfect agreement
between the numerical simulation results and the exact solution for $E_{0a}$
as shown in Fig. 1.

Extending this method to the four dimensional Ising model we confronted
our numerical results with two theoretical predictions, one coming from
a dilute instanton gas approximation \cite{munster}, eq.(\ref{eq:E0a})
and one recent work, claiming to have obtained the exact finite size behavior
for $E_{0a}$ \cite{borgs}, eq.(\ref{eq:bie0a}).
We measured the infinite volume values of
$m_r$ and $g_r$
(as they are required as input for the instanton calculation)
at four different $\kappa$ values (see Table 2).
The measured values of $E_{0a}$ (see Table 1)
agree reasonably with the predictions by eqs.
(\ref{eq:E0a}), (\ref{eq:sigmainstanton}) and (\ref{eq:cinstanton}) from
the dilute instanton gas approximation as shown in Fig. 2. Our fits to the
form of
eq. (\ref{eq:fit1}) or eq. (\ref{eq:fit2}) give comparable $\chi^2$ per
degrees of freedom. This suggests that our numerical data are not accurate
enough to determine the power of the $L$ dependence
in the pre-exponential factor of $E_{0a}$.
However, assuming $E_{0a}$ has the form of eq. (\ref{eq:E0a}) with
$\omega=(d-3)/2$, the fitted
values for $\sigma$ and $C$ agree well with the predictions given
by eqs. (\ref{eq:sigmainstanton}) and (\ref{eq:cinstanton}) as shown in
Table 3. In contrast, a fit to the form eq. (\ref{eq:fit2}), shown in
Table 4,  gives $C$ values that are much smaller than two and $\kappa$
dependent. This is in contradiction to the result of Borgs and
Imbrie, eq. (\ref{eq:bie0a}).

We conclude that given the choice between the instanton calculation
and the result of ref. \cite {borgs}, our numerical data
favor the instanton calculation result. Two comments are in order.
First our numerical simulations are performed close to $\kappa_c$
so that the system is inside the scaling region. The instanton calculation
is only valid inside the scaling region since the domain walls are treated as
continuous surfaces and renormalization language is used. Second,
at present we can not determine the value of $\omega$ in eq. (\ref{eq:E0a})
from our numerical data. Hypothetically,
it is possible to have a scenario that $E_{0a}$ takes the form in
eq. (\ref{eq:fit2}) with a $\kappa$ dependent $C$. Close to the critical
point, $C$ would be much smaller than two and gradually increases to two
when $\kappa$ becomes large. We emphasize that so far this scenario
has no theoretical backing. In order to determine the value of $\omega$
further numerical simulation is needed on a larger lattice and with better
statistics.

\noindent{Acknowledgement}

We thank M. Creutz, J. Z. Imbrie, J. Kuti and G. M{\" u}nster for useful
discussions.
Our work is supported by DOE grants at UC San Diego (DE-FG-03-90ER40546)
and Brookhaven National Laboratory (DE-AC02-76CH00016). The numerical
simulations are performed at UCSD Supercomputer Center and SCRI at Tallahassee.

\pagebreak

\section*{Table Caption}

\noindent {\bf Table 1}: We give the values of the energy splitting
$E_{0a}$ as
determined from fits to the correlation function
eq. \protect{(\ref{eq:correlat})}.
The lattice geometry has been $TL^3$ where $T=120$ for $\kappa =0.0760$
and $T=160$ elsewhere.
The values at $\kappa =0.0760$ (except the L=11 result) are from
reference \protect{\cite{jansen}}.

{\bf Table 2}: We give the renormalized mass $m_r$ and the renormalized
coupling $g_r$ for the four dimensional Ising model for all $\kappa$ values
used. The values for $\kappa =0.0760$ are taken from reference
\protect{\cite{jansen}}.

\noindent {\bf Table 3}: We take $E_{0a}$ measured on the largest three
volumes at each $\kappa$ value and fit the data according to eq.
(\ref{eq:fit1}). The fitted $\sigma_{fit}$ and $C_{fit}$ are listed here
along with $\chi^2$ per degrees of freedom. For comparison $\sigma_{inst}$
and $C_{inst}$ calculated from eqs. (\ref{eq:sigmainstanton})
and (\ref{eq:cinstanton}) are also listed.

\noindent {\bf Table 4}: Similar to Table 3, but data are fitted
to the form of eq. (\ref{eq:fit2}). Note $C$ is much smaller than two
and dependent on $\kappa$.

\section*{Figure Caption}

{\bf Figure 1}: The energy splitting $E_{0a}$ in the two
dimensional Ising model as obtained from the Monte Carlo simulation. The
open symbols are at $\kappa =0.2326$, the solid symbols are at $\kappa
=0.2252$. The errorbars are smaller than the size of symbols.
The lines represent the exact solution
eq. \protect{(\ref{eq:exact1})}. Note that no fit is involved.

\noindent {\bf Figure 2}: The energy splitting $E_{0a}$ for the four
dimensional Ising model at four $\kappa$ values. The solid circles
indicate the measured $E_{0a}$. The statistical errors are smaller
than the size of symbols. The full curves
are the {\em predictions} from the instanton calculation eq. (\ref{eq:E0a})
\protect{\cite{munster}}.

\noindent {\bf Figure 3}: The energy splitting $E_{0a}$ for the four
dimensional Ising model at $\kappa=0.0754$. The dashed curve is given by
eq. (\ref{eq:bie0a}) with $\sigma$ calculated from
eq. (\ref{eq:sigmainstanton}). For comparison the full curve
is from the instanton calculation eq. (\ref{eq:E0a}).

\pagebreak

\begin{center}

\begin{tabular}{||r|l|r|l|r|l|r|l||}\hline
\multicolumn{2}{||c|}{$\kappa =0.0751$} &
\multicolumn{2}{|c|}{$\kappa =0.0754$} &
\multicolumn{2}{|c|}{$\kappa =0.0760$} &
\multicolumn{2}{|c||}{$\kappa =0.0764$} \\ \hline
$L$  & $E_{0a}$ & $L$ & $E_{0a}$   & $L$ & $E_{0a}$ & $L$ & $E_{0a}$ \\ \hline
 7   & 0.161(1)  &  7  & 0.1345(8) &  6  & 0.1281(4)&  5 & 0.181(2)  \\ \hline
 8   & 0.1320(8) &  8  & 0.1015(8) &  7  & 0.0812(3)&  6 & 0.098(2)  \\ \hline
 9   & 0.1118(7) &  9  & 0.0761(8) &  8  & 0.04609(2)& 7 & 0.0533(8) \\ \hline
 10  & 0.0944(7) & 10  & 0.0552(9) &  9  & 0.02238(1)& 8 & 0.0223(3) \\ \hline
 11  & 0.0757(8) & 11  & 0.0370(6) & 10  & 0.00902(6)& 9 & 0.00726(4)\\ \hline
 12  & 0.0609(7) & 12  & 0.0270(11)& 11  & 0.0027(3) & 10& 0.00190(6)\\ \hline
 13  & 0.0509(7) & 13  &           &     &           &   &           \\ \hline
 14  & 0.0436(15)& 14  &           &     &           &   &           \\ \hline
\end{tabular}

\vspace{0.6cm}
{\bf Table 1}

\vspace{2.0cm}

\begin{tabular}{||l|l|l||} \hline
$\kappa $ & $m_r$      & $g_r$   \\ \hline
 0.0751 &  0.169(3)  &  24.0(9)  \\ \hline
 0.0754 &  0.264(1)  &  28.3(6)  \\ \hline
 0.0760 &  0.395(1)  &  30.2(4)   \\ \hline
 0.0764 &  0.484(2)  &  36.3(4)   \\ \hline
\end{tabular}

\vspace{0.6cm}
{\bf Table 2}

\vspace{2.0cm}

\begin{tabular}{||l|l|l|l|l|l||} \hline
$\kappa $ & $\sigma_{fit}$ & $\sigma_{inst}$  & $C_{fit}$ & $C_{inst}$&
$\chi^2$  \\ \hline
 0.0751 &  0.00043(3)  &0.00037(2) & 0.037(2) & 0.033(1) & 0.6  \\ \hline
 0.0754 &  0.00121(5)  &0.00117(3) & 0.057(3) & 0.0595(7)& 3.2  \\ \hline
 0.0760 &  0.00358(2)  &0.00363(6) & 0.101(1) & 0.1054(9)& 0.4   \\ \hline
 0.0764 &  0.00536(6)  &0.00543(9) & 0.121(5) & 0.130(1) & 1.6  \\ \hline
\end{tabular}

\vspace{0.6cm}
{\bf Table 3}

\vspace{2.0cm}
\begin{tabular}{||l|l|l|l||} \hline
$\kappa $ & $\sigma$ fitted  & $C$ fitted & $\chi^2$  \\ \hline
 0.0751 &  0.00035(3)  &0.112(7) & 0.4  \\ \hline
 0.0754 &  0.00107(5)  &0.16(1)  & 2.7  \\ \hline
 0.0760 &  0.00334(2)  &0.255(2) & 0.1   \\ \hline
 0.0764 &  0.00511(6)  &0.30(1)  & 0.9  \\ \hline
\end{tabular}

\vspace{0.6cm}
{\bf Table 4}

\end{center}


\begin{thebibliography}{99}

\bibitem{fisher} M.E. Fisher, J. Phys. Soc. Japan Suppl. 26 (1969) 87.

\bibitem{munster} G. M\"unster, Nucl. Phys B324 (1989) 630; Nucl. Phys.
B340 (1990) 559.

\bibitem{brezin} E. Brezin and J. Zinn-Justin, Nucl. Phys. B257 [FS14]
(1985) 867.

\bibitem{jansen} K. Jansen, J. Jers\'ak, I. Montvay, G. M\"unster, T.
Trappenberg and U. Wolff, Phys. Lett. B213 (1988) 203;\\
               K. Jansen, I. Montvay, G. M\"unster, T.
Trappenberg and U. Wolff, Nucl. Phys. B322 (1989) 698;\\
K. Jansen, Nucl. Phys B (Proc. Suppl.) 9 (1988) 35.

\bibitem{mot}H. Meyer-Ortmanns and T. Trappenberg, J. Stat. Phys. 58
(1990) 185.

\bibitem{borgs} C. Borgs, J.Z. Imbrie, Harvard University preprint,
91-0233.

\bibitem{onsager} L. Onsager, Phys. Rev. 65 (1944) 117.

\bibitem{kushe} J. Kuti and Y. Shen, Phys. Rev. Lett. 60 (1988) 85.

\bibitem{luwe} M. L\"uscher and P. Weisz, Nucl. Phys. B295 [FS21] (1988)
65.

\bibitem{privman} V. Privman and M.E. Fisher, J. Stat. Phys. 33 (1983)
385.

\bibitem{coleman} S. Coleman, Aspects of Symmetry (Cambridge University
Press, Cambridge, 1985).

\bibitem{swendsen} R.H. Swendsen and J.-S. Wang, Phys. Rev. Lett. 58
(1987) 86.

\bibitem{wolff} U. Wolff, Phys. Rev. Lett. 62 (1989) 361; Nucl. Phys. B
(Proc. Suppl.) 17 (1990) 93.

\bibitem{sweeny} M. Sweeny, Phys. Rev. B27 (1983) 4445.

\end{thebibliography}
\end{document}